# Experimental Outdoor Performance Evaluation of TVWS Narrowband Data Communication using SDR Platform

Muneer M. AlZubi, *Member, IEEE*, and Mohamed-Slim Alouini, *Fellow, IEEE*

*Abstract*— Large-scale deployment of Internet of Things (IoT) networks in the industrial, scientific, and medical (ISM) band leads to spectrum congestion and requires multiple gateways to cover wide areas. This will increase cost, complexity, and energy consumption. TV White Spaces (TVWS) provides an abundant spectrum that is sufficient for low data rate IoT applications. This low-frequency band offers coverage over larger areas due to the ability of wireless signals to penetrate obstacles and terrain. In this paper, we examine the performance of narrowband data communications in TVWS through an outdoor experiment in a suburban area with line-of-sight (LOS) and non-line-of-sight (NLOS) propagation scenarios. We implement a software-defined radio (SDR) testbed and develop a GNU radio benchmark tool to perform outdoor experiments for TVWS narrowband data communication between a gateway and wireless nodes at various locations. The results reveal that the system can achieve a throughput of up to 97 Kbps with a packet error rate (PER) and packet loss rate (PLR) under 1% over NLOS paths, making it suitable for low-data rate applications. This work offers valuable insights for designing the physical layer of narrowband white space devices (WSDs). The developed benchmark tool will also greatly assist other researchers in evaluating the performance of SDR-based communication systems.

## I. INTRODUCTION

THE Internet of Things (IoT) has been increasingly widespread around the world in recent years, and hundreds of billions of IoT devices are expected to be in use in the next few years [1, 2]. IoT networks are used in different applications and fields such as agriculture, healthcare, manufacturing, transportation, environmental monitoring, smart cities, and smart grids. There are various wireless communication solutions for IoT networks, including short-range and long-range technologies [3]. The short-range technologies, aka wireless personal area networks (WPANs), provide short communication distance with either high or low data rates and thus with higher or lower power consumption, such as Wi-Fi, Bluetooth, Zigbee, and Z-Wave. On the other hand, low power wide area networks (LPWANs) provide a longer communication range with low power consumption but with low data rates using licensed or unlicensed frequency such as LoRa WAN, Sigfox, Weightless, and narrowband IoT (NB-IoT). LPWANs suit IoT applications that require small data packets to be transmitted infrequently or periodically, e.g., hourly or daily, over long distances with low power consumption.

The rapid increase in IoT deployment leads to spectrum congestion, particularly in the industrial, scientific, and medical (ISM) bands with limited bandwidth. This could impact IoT applications that require a reasonably high to moderate data rate to speed up the transfer of large amounts of data, such as camera images, real-time data, and firmware. Moreover, current IoT deployment at high frequency in the ISM band requires multiple gateways to cover large areas, which increases power, cost and complexity. This is due to the short propagation distance that a high frequency radio signal can reach compared to a low frequency signal. This problem can particularly affect IoT applications in rural and remote regions that cover wide areas such as solar/wind farms, gas/oil sites, and remote towns. The lack of infrastructure and low population density in these areas require cost-effective solutions to provide wide coverage at a reasonable data rate for IoT networks. A promising solution to these challenges is using narrowband data communications in the TV White Space (TVWS) band. TVWS refers to unused channels in the television broadcast spectrum by licensed services in very high frequency (VHF) and ultra-high frequency (UHF) bands that can be used for broadband Internet access and IoT applications. TVWS provides a solution to the common challenges of spectrum congestion in licensed and unlicensed ISM bands. TVWS offers an abundant spectrum for low to high data-rate applications (e.g., 470-790 MHz in the UHF band). In addition, compared to the higher frequency signals used in some technologies, such as Wi-Fi and Bluetooth, the TVWS radio signal can easily penetrate obstacles, terrain and vegetation, providing longer transmission ranges. Therefore, TVWS can provide a longer communication range for many IoT devices using star topology, which reduces the number of gateways and thus reduces cost and energy.

So far, TVWS is primarily exploited by industry leaders, such as 6Harmonics and Adaptrum, to provide broadband internet connectivity in many regions worldwide [4-6]. However, no commercial narrowband white space (WSD) devices have been manufactured yet. This could be due to several reasons,

The authors are with the Computer, Electrical and Mathematical Science and Engineering Division (CEMSE), King Abdullah University of Science and Technology (KAUST), Thuwal 23955-6900, Makkah Province, Saudi Arabia (e-mail: muneer.zubi@kaust.edu.sa; slim.alouini@kaust.edu.sa).

such as the delay in adopting narrowband TVWS systems by regulatory bodies, which was first announced by the Federal Communications Commission (FCC) in 2020 [7]. Another reason is the low power requirements and potential harmful interference of WSDs to primary services such as TV receivers and wireless microphones, which impose challenges and costs on design requirements such as spectrum sensing techniques and communication protocols with the geolocation database. However, TVWS is the ideal solution for narrowband IoT applications that require extended coverage, low to medium data rates, and low power consumption. It is worth mentioning here that the concept of the narrowband communication system proposed in TVWS is different from the narrowband Internet of Things (NB-IoT) induced under 3GPP cellular technology.

Few studies considered narrowband data transmission in the TVWS band. In [8], a radio access mechanism is proposed for massive IoT over TVWS based on regulatory policies for interacting with the geolocation database via the protocol to access white space (PAWS). In [9], an experimental testbed using Gaussian minimum-shift keying (GMSK) was implemented to test the TVWS narrowband transmission inside a laboratory. In [10], a transceiver prototype for the wireless smart utility network (Wi-SUN) using TVWS and IEEE 802.15.4m standard [11, 12] is developed to improve coverage, bandwidth, and scalability. The IEEE 802.15.4m wireless standard is proposed to enable low-data-rate WPANs operating in the TVWS band. The performance of the prototype developed using direct cable connection between transceivers is examined in terms of bit error rate (BER) using the forward error correction (FEC) technique and phase-shift keying (PSK) modulation scheme. A modified IEEE 802.15.4-based LPWAN architecture is developed using software-defined radio (SDR) by exploiting the TVWS using binary PSK (BPSK) modulation technique over a channel bandwidth of 200 kHz to 1 MHz [13, 14]. However, the studies mentioned above were mostly performed inside the laboratory with a wireless line-of-sight (LOS) link or a direct wire cable connection without considering the realistic outdoor propagation channel. Another limitation is that they did not evaluate the performance considering the regulation rules for narrowband transmission in the TVWS spectrum. In addition, these studies did not provide performance comparison using various modulation techniques under the same channel conditions and experiment parameters. Recently, two studies have focused on developing LoRa-based wireless communication systems in TVWS for IoT applications [15, 16]. In [15], the authors studied LoRa transmissions in the TVWS band through theoretical analysis to examine data rate and coverage range. In [16], a narrowband LoRa-based TVWS wireless communication system is developed and examined following the regulation rules for operating narrowband communications in the TVWS spectrum. However, LoRa relies on spread spectrum modulation which poses challenges in designing spectrum sensing technologies compared to traditional methods such as energy detection (ED) which do not need prior knowledge about the primary user signal. Spectrum sensing is critical to cognitive radio (CR) technology, such as the TVWS systems. The proposed sensing technique for LoRa signal is limited to the type of wireless devices that use LoRa modulation, which does not provide scalability and compatibility with other WSDs operating in the TVWS systems such as wireless microphones. Moreover, embedding this sensing technique into WSDs makes narrowband devices more complex and expensive. The ED technique provides several advantages in terms of application and computation complexities and can be used with various physical layers and communication technologies.

In addition, since we focus here on developing a narrowband SDR-based wireless communication system in the TVWS band using GNU Radio, we need to evaluate the over-the-air system performance in terms of various performance metrics using reliable and flexible benchmark tools. The GNU Radio is a free software development toolkit that provides the capability to design SDR-based communication systems using either external radio hardware or simulation platforms [17]. Since the release of GNU Radio software, the only available benchmark tool used by the GNU Radio community was a simple built-in component in the previous GNU Radio versions [18]. However, this tool has been removed from recent GNU Radio versions. It was used via a command-line interface, passing specific input parameters such as modulation type, frequency, TX/RX gain, and packet size. In this tool, the benchmark_tx.py file generates and modulates the data packets and sends them across the air via a universal software radio peripheral (USRP) device while the benchmark_rx.py file receives these packets and provides the error rate. Over the past two decades, various studies have used the same benchmark files or modified versions to evaluate the system performance, e.g., [19-24]However, this tool has been removed from recent releases due to its inefficiency and unreliability. For example, some researchers have found that it drops data packets unexpectedly for unknown reasons. Furthermore, it did not provide flexibility in supporting different SDR devices and communication systems. Therefore, the data transmission systems developed using this tool are restricted by its capability and requirements. Moreover, it did not support the GUI GRC flowgraph in GNU Radio. Finally, it did not provide information on other performance metrics such as latency, signal-to-noise ratio (SNR), and packet loss rate (PLR). Hence, there is a great need for a reliable and flexible benchmark tool capable of solving the limitations mentioned above.

In this work, we develop an end-to-end SDR-based TVWS narrowband data communication system using various modulation schemes considering the TVWS regulation rules, such as the maximum allowed effective isotropic radiated power (EIRP), channel bandwidth, and antenna gain. Moreover, we built a flexible and easy-to-use benchmark module for evaluating the performance of the developed communication system. It is a generalized module that can be used with various SDR-based communication systems and modulation techniques in GNU Radio. Using the developed testbed and benchmark tool, we conducted outdoor over-the-air measurements in suburban areas on the King Abdullah University of Science and Technology (KAUST) campus. In this measurement campaign, we consider reasonable and realistic scenarios, such as choosing reasonable antenna heights for the gateway and the nodes and examining non-line-of-sight (NLOS) radio propagation links. We evaluate the system performance using the developed benchmark module in terms of throughput, latency, packet error rate (PER), PLR, SNR, and distance. This work may provide insight to help develop narrowband TVWS communication systems using efficient modulation techniques in future work. Also, the developed benchmark module can help the GNU Radio community evaluate the performance of various SDR-based communication systems.

The rest of the paper is organized as follows. In Section II, an overview of the TVWS system and its regulations are presented. Section III presents the system architecture of the developed testbed including the benchmark module and the outdoor measurement experiment. The performance results are discussed in Section IV. The paper is concluded in Section V.

## II. OVERVIEW OF TVWS TECHNOLOGY

In general, white space refers to unused frequency gaps in the licensed spectrum. TVWS refers to free channels in the VHF and UHF TV bands made available for unlicensed use by secondary systems at locations where licensed primary systems do not use the spectrum. Primary systems include digital terrestrial television (DTT) and programme-making and special events (PMSE) [25]. For example, the TVWS spectrum ranges from 470-790 MHz in Europe and 54-698 MHz in the United States [26]. Many regulatory regimes around the world have authorized unlicensed access to the TVWS spectrum for wireless communication services. TVWS is superior to other technologies operating at higher frequencies, such as Wi-Fi and long-term evolution (LTE). It can provide Internet connectivity with a longer transmission range and better penetration through terrain, trees, and buildings which can help bridge the digital divide. TVWS can be seen as an important practical application of cognitive radio and spectrum-sharing technology via dynamic spectrum access. In TVWS, the regulatory authorities allow WSDs to broadcast on these free channels without a license, provided that they are registered in a specialized geolocation database to protect licensed users from harmful interference., see Fig. 1. The WSDs should contact the base station (BS) to obtain a list of available channels from the geolocation database and the corresponding maximum power levels that can be used at their locations. Therefore, WSDs do not transmit on the channels used by primary TV broadcasters within a specific area and thus will not cause interference to primary users. In general, secondary users allowed to use the TVWS band are referred to as WSDs where the BS is referred to as master WSD and the client is referred to as slave WSD or customer premises equipment (CPE). In TVWS, WSDs can be either fixed or portable, where fixed users are usually allowed to transmit at higher power than portable users. Furthermore, WSDs can operate either in an independent mode requiring database access to obtain available channels or in client mode without database access. The TVWS spectrum is abundant in both VHF and UHF bands. In the last decade, TVWS has been widely used to provide broadband Internet connectivity to rural and hard-to-reach regions in many countries using several commercial TVWS equipment manufactured by leading vendors such as 6Harmonics and Adaptrum [4, 27-29]. In addition, TVWS can be used to increase the coverage area and cover missing spots in other wireless technologies such as 5G/6G and citizens broadband radio service (CBRS) networks.

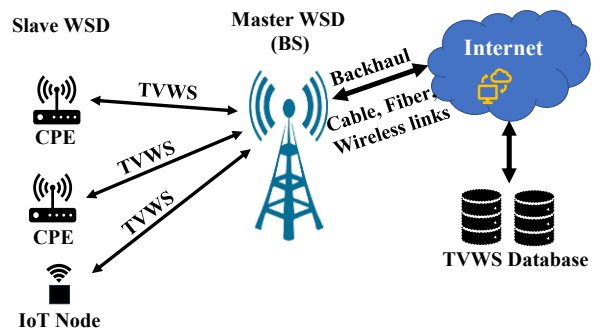

**Fig. 1.** TVWS system architecture.

Recently, the FCC adopted regulation rules to facilitate the development of new and innovative narrowband WSDs (NB-WSDs) operating in the TVWS band for IoT applications [7]. The current rules define the NB-WSD as a fixed or portable WSD operating in a channel bandwidth of no more than 100 KHz. The NB-WSD can be either a client or master device where the client communicates with the master device that contacts the white space database to obtain a list of available channels and operating powers at its location. Under current FCC rules, fixed devices should have a maximum antenna gain of 6 dBi with no reduction in transmitter-conducted power or higher antenna gain if the conducted power is proportionally reduced. However, if transmitting antennas of directional gain greater than 6 dBi are used, the maximum conducted output power should be reduced by an amount in dB that the directional antenna gain exceeds 6 dBi. Moreover, fixed WSDs operating with EIRP larger than 4W (36 dBm) must comply with a power spectral density (PSD) limit of 12.6 dBm/100kHz (i.e., EIRP limit of 18.6 dBm/100 kHz) during any time interval of continuous transmission. This will limit total conducted power within any 6 MHz TV channel to 30 dBm. Also, fixed devices should comply with an emission limit of -42.8 dBm in 100 kHz into adjacent channels, i.e., outside of the 6 MHz channel on which they operate. The NB-WSD follows the same rules as fixed WSD regarding the conducted PSD limit, adjacent channel emission limits, and antenna gain requirements. Each NB-WSD can transmit for no more than 36 seconds/hour (i.e., a 1% duty cycle) to ensure that the total energy in a single TV channel does not cause harmful interference and prevent NB-WSDs from being used for data-intensive applications.

In addition, NB-WSDs should use a channel plan that limits total transmitted power in a 6 MHz channel to no more than the existing limits for a 4W EIRP broadband WSD. This channel plan requires NB-WSDs to operate at least 250 kHz from the upper and lower edges of a 6 MHz TV channel unless the adjacent channel is also vacant or in the case of bonded 6 MHz channels sharing a common band edge, then no offset is needed. Also, NB-WSDs should operate only on channels centered at integral multiples of 100 kHz between the 250 kHz guard bands. Thus, NB-WSDs can operate within 55 possible 100 kHz subchannels from the center of each 6 MHz channel (i.e., 5.5 MHz). Even if all 55 narrowband subchannels within a 6 MHz channel were occupied simultaneously by NB-WSD transmitting at maximum power, the total conducted and radiated power within that 6 MHz channel would be no greater than for a fixed WSD operating with conducted power of 1W (30 dBm) and EIRP of 4W (36 dBm). Moreover, since the transmission time is limited to 1% duty cycle, the interference potential of these NB-WSDs will be significantly less than the fixed devices in most cases since it is extremely unlikely that NB-WSDs would transmit at maximum power on all 55 narrowband subchannels simultaneously, and even if they did, that would occur for no more than 36 seconds per hour. These rules will allow NB-WSD to operate with single or several narrowband carriers while ensuring that NB-WSDs have no greater interference potential than broadband WSD operating under the current rules. Finally, it is worth mentioning that the FCC restricts the TV frequency that can be used for N-WSDs to less than 602 MHz.

## III. SYSTEM ARCHITECTURE

This section describes the developed SDR-based narrowband TVWS testbed, including hardware and software parts implemented using GNU Radio and USRP. We also present the benchmark module and the outdoor measurement campaign for collecting data and performance metrics.

### A. Experimental Measurement Testbed

In this subsection, we demonstrate the radio measurement testbed used for collecting channel measurements and system performance of narrowband wireless communication in the TVWS band. The hardware testbed is implemented using two USRP SDR devices, one acts as a BS and the other acts as a node, which are connected to omnidirectional monopole antennas as shown in Fig. 2. The USRP devices are connected to laptops running the GRC flowgraph files developed in GNU Radio as will be described in the next subsections. Two portable power stations were used to power the USRPs and laptops.

TABLE I
THE MEASURED POWER CALIBRATION DATA.

| Normalized Gain | Channel Power / 100kHz (dBm) | PSD (dBm/Hz) | Peak Magnitude (dBm) |
|---|---|---|---|
| 0 | -11 | -61 | -38 |
| 0.1 | -8.2 | -58.2 | -36 |
| 0.2 | -5.4 | -55.4 | -33.7 |
| 0.3 | -2.4 | -52.4 | -29.8 |
| 0.4 | 1 | -49 | -27 |
| 0.5 | 3.8 | -46.2 | -24.2 |
| 0.6 | 6.8 | -43.2 | -21.5 |
| 0.7 | 10.2 | -39.8 | -17.7 |
| 0.8 | 13.2 | -36.8 | -16.3 |
| 0.9 | 16 | -34 | -11.8 |
| 1 | 18.2 | -31.7 | -8.6 |

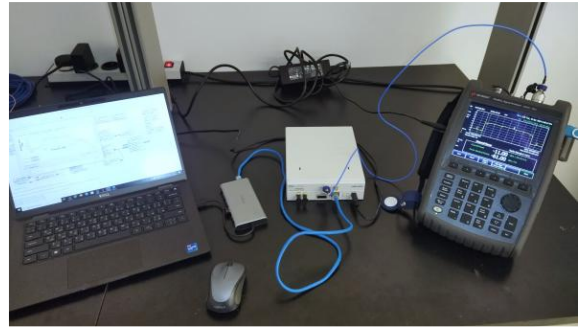

**Fig. 3**. Power calibration setup in the laboratory using USRP N210 and Keysight-FieldFox spectrum analyzer.

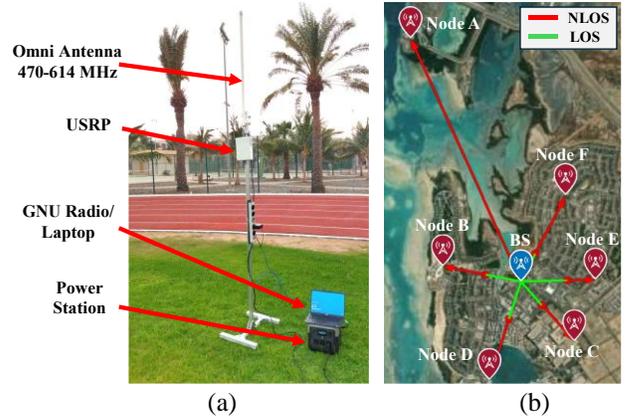

(a)                    (b)

**Fig. 4**. The outdoor measurement setup: (a) the node testbed includes the USRP, power station, laptop, and monopole antenna, and (b) the locations of nodes and BS on the KAUST campus.

It is worth mentioning here that the USRP is not a calibrated device. Therefore, it does not provide the ability to adjust the absolute transmitted power at the transmitter and to measure the absolute received signal strength (RSS) at the receiver. However, it uses relative power level rather than absolute value. Therefore, we first perform a calibration experiment in the laboratory before using the USRP devices. The relative transmitted power in the USRP can be controlled by adjusting the gain using either a normalized value between 0 and 1 or a power value in dB or dBm. We use a normalized value between 0 and 1 to change the transmitted power where 0 and 1 represent the minimum and maximum power levels, respectively. The transmitter USRP is connected to the Keysight-FieldFox spectrum analyzer via a coaxial cable, as shown in Fig. 3. Then, the GRC block diagram of the transmitter shown in Fig. 5 is executed. We vary the gain value between 0 and 1 and then record the corresponding measured values of the channel power, the power spectral density (PSD), and the peak magnitude, as listed in Table I. To comply with the maximum EIRP allowed in FCC regulations (i.e., 18.6 dBm), we adjust the normalized gain to a value less than 0.8 with an omnidirectional antenna having a gain of 5 dBi.

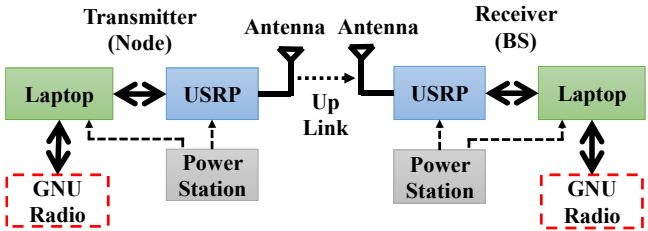

**Fig. 2.** An illustrative block diagram of the radio measurement testbed.

TABLE II
THE NODES AND BS LOCATIONS ON THE KAUST SUBURBAN CAMPUS.

| Node | Location | Latitude | Longitude | Distance (m) |
|------|----------|----------|-----------|--------------|
| A | King Abdullah Monument | 22.34142474 | 39.08886671 | 3,122 |
| B | Island Recreation Club (IRC) | 22.31739426 | 39.09256216 | 935 |
| C | Building 9 | 22.30946328 | 39.10747198 | 933 |
| D | Al Marsa Restaurant | 22.30536219 | 39.09786981 | 1,212 |
| E | Nargis Lane Park | 22.31597395 | 39.10989160 | 867 |
| F | Skateboard Park | 22.32484347 | 39.10647471 | 1,128 |
| BS | Garden | 22.31578185 | 39.10147449 | - |

Before conducting the outdoor measurements, we obtained permission from the Communications, Space and Technology Commission of Saudi Arabia to use the TV spectrum within the KAUST campus. We have conducted outdoor field measurements on the KAUST campus using the developed SDR-based testbed, see Fig. 4. The KAUST campus is a suburban area that includes both university buildings and residential areas. The residential areas consist mostly of two-storey townhouses, detached houses and villas. The transmitter (node) is placed at various locations with an antenna mounted on a 2m mast as shown in Table II and Fig. 4. The BS is installed on a house's roof with an antenna height of 12m above ground level (i.e., the mast length is about 2m). It is worth mentioning here that we use a reasonable antenna height for the BS, which can be easily deployed in suburban and rural areas. Most IoT networks have a star topology where each node connects directly to common central access points or gateways. Therefore, we focus on the communication links between the nodes and the BS as in the star topology. The laptops are remotely connected over the Internet to facilitate remote control of configuration parameters on the BS side. The outdoor measurements are performed for various modulation schemes, including DBPSK, DQPSK, GFSK, and

GMSK. In addition, the performance metrics are measured and analyzed under different transmit power levels. We chose these modulation schemes because they are commonly used in digital communication systems that require low data rates and/or low power consumption, making them effective techniques for IoT applications.

## B. System Development in GNU Radio

In this subsection, we describe the block diagram of the developed communication system using the GNU Radio software. The developed system is executed on the Ettus Research™ USRP N210 SDR platform. We are more concerned with the uplink connection in IoT applications, where most data are transferred from the IoT device to the gateway (BS). Therefore, this system consists of an IoT node that acts as a transmitter and a BS that acts as a receiver. Figure 5 shows the transmitter's block diagram for various modulation techniques. The first block is the data source (file source) which represents the data that needs to be transmitted, such as text, audio waveform, image, etc. In this system, we use a text file of size 364 KB as a data source. The data is fed into the cyclic redundancy check (CRC) generator block, which generates and appends a 32-bit CRC code to the payload. This output represents the payload that will be transmitted. The packet formatter block includes a protocol formatter block that generates the headers and then combines the payloads and headers to create the packets. Now, the generated packets represent the baseband signal ready to be modulated using different modulation schemes, including DBPSK, DQPSK, GFSK, and GMSK [30].

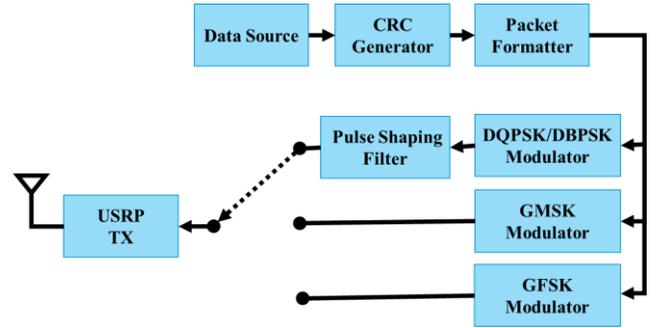

**Fig. 5**. The transmitter (i.e., node) block diagram.

In differential PSK (DPSK) modulation, the phase difference between consecutive symbols is used to represent digital data rather than the absolute phase, as with ordinary PSK. For example, in DBPSK, the binary bits "1" and "0" can be transmitted by adding 180° and 0° to the current phase, respectively. In DQPSK, the data symbols "00", "01", "11", and "10" can be transmitted using phase shifts of 0°, 90°, 180°, and -90°, respectively. The DPSK can be implemented by adding a deferential encoder at the PSK modulator input and a differential decoder at the PSK demodulator output. As an alternative to ordinary PSK, DPSK eliminates the need for a complex carrier-recovery scheme to estimate the absolute phase. The DPSK is used in communication channels where robustness against phase variations is critical. Therefore, the DPSK is less susceptible to absolute phase variations caused by the wireless channel impairments, and it requires a simpler phase tracking system at the receiver compared to the ordinary PSK. The communication channel generally causes unknown phase shifts to the transmitted signal and thus, DPSK can produce a lower bit rate in such a channel than the ordinary PSK. The DPSK-modulated signal is fed into a root-raised cosine (RRC) pulse-shaping filter that controls the bandwidth of the transmit signal to prevent power leakage and interference with the adjacent channels. In the binary FSK (BFSK), the transmitted "0" and "1" binary bits are modulated using two distinct frequencies. However, the process of instantaneously switching to a different frequency in practice may lead to phase discontinuity and sudden amplitude variation. This increases the sideband power which causes interference to neighboring channels. In the GFSK modulation technique, a Gaussian filter is used at the input of the FSK modulator to make the transition between discrete frequencies smoother to reduce the spectral width, i.e., it acts as a pulse-shaping filter in this application. The Minimum-shift keying (MSK) is a modulation scheme that can be viewed as a continuous phase FSK (CPFSK) with a frequency separation of one-half the bit rate. In other words, the waveforms used to represent the bits "0" and "1" differ by half the carrier period. Accordingly, the modulation index m is equal to ½, which is the smallest FSK modulation index that can be chosen so that the waveforms for bits "0" and "1" are orthogonal. This leads to a constant envelope signal that reduces problems caused by non-linear distortion. In Gaussian MSK (GMSK), the digital data is first shaped using a Gaussian filter before the standard MSK modulator to reduce the sideband power which decreases the out-of-band interference in adjacent channels. By reducing the out-of-band emissions, we keep the transmitted signal within the channel bandwidth of 100 KHz. Then, the modulated signal is fed into the USRP TX (USRP Sink) block that acts as a radio transmitter hardware. The advantage of DBPSK is that it requires a simpler hardware design and is less sensitive to phase variations compared to DQPSK. However, DQPSK is more complex but provides a higher data rate compared to other modulation schemes. DPSK modulation technique carries data more efficiently, i.e., more efficient use of bandwidth, with less power consumption compared to GFSK. However, DPSK requires complex detection and recovery systems compared to other techniques. Compared to GFSK, GMSK minimizes the bandwidth while maintaining a constant envelope, simplifying power amplification and making it suitable for power-efficient and non-linear power amplifiers.

On the BS side, we have implemented the receiver, which operates with the four modulation schemes mentioned above. The first part of the receiver is the USRP RX (USRP Source) block which acts as radio receiver hardware. A sample of the received signal spectrum from the USRP RX block is shown in Fig. 6 (a). We can observe that the frequency spectrum of the received signal is almost limited to 100 kHz channel bandwidth. First, we have developed the GFSK and GMSK receivers, as shown in Fig. 7. The received signal passes through an automatic gain control (AGC) block that adjusts the input signal to a stable reference level suitable for demodulation. The output signal is then filtered using a low pass filter (LPF) that keeps the desired frequency part of the signal and removes the unwanted noise, see Fig. 6 (b). Then, the complex signal is demodulated using either a GFSK or GMSK demodulator. The output data from the demodulator is a stream of bits representing the received packets. The extract payload block includes a correlate access code block that searches in the bit stream for the packet header to get the packet length and then the payload bits. The payload data is then checked using a CRC check block to check whether the CRC is correct or incorrect. If the CRC is correct, the payload data will be exported to an external file using the data output block (file sink block).

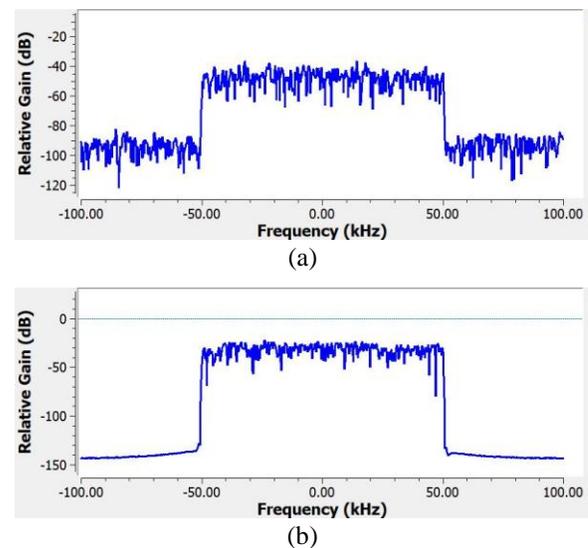

**Fig. 6**. A sample of the frequency spectrum of a real-time narrowband signal received by the BS: (a) the original received signal and (b) the filtered signal using the LPF.

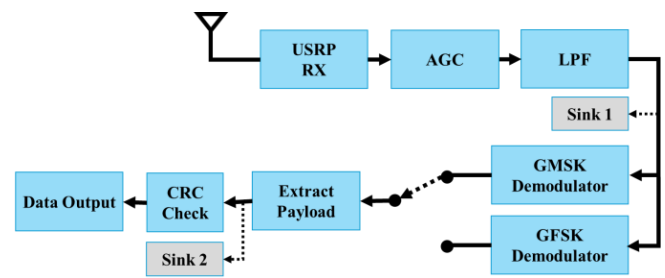

**Fig. 7**. The GFSK and GMSK receiver block diagram.

The block diagram of the DBPSK/DQPSK receiver is shown in Fig 8. As in the GFSK/GMSK receiver, the output samples from the USRP RX block are passed into the AGC and the LPF blocks. Then, the filtered signal is fed into the Symbol Synchronization block that performs clock recovery to synchronize with the symbols in the digital signal. In this system, we used a pulse-shaping RRC filter in the transmitter, which may create inter-symbol interference (ISI). Therefore, another RRC-matched filter is needed at the receiver to minimize the ISI. Now, the stream of complex samples is equalized using the Linear Equalizer block to mitigate the signal distortion and attenuation caused by the multipath fading. The equalizer is provided with an adaptive constant modulus algorithm (CMA) to equalize the channel. However, the signal still suffers from phase and frequency offset problems that need to be corrected. Thus, the Carrier Recovery (Costas loop) block uses a second-order phase-locked loop (PLL) to track both phase and frequency to synchronize both DBPSK and DQPSK signals. The Costas loop locks to the signal's centre frequency and down-converts it to the baseband. Then, the output is passed into the DBPSK/DQPSK demodulator block, which includes both the constellation decoder and differential decoder blocks. Thus, it first decodes the constellation's points from the complex space to binary bits for either QPSK or BPSK. Then, it performs differential decoding to transform the differentially coded symbols back to their original transmitted symbols based on the phase transitions of the current and previous symbols. It is worth mentioning that the ambiguity in knowing the transmitted symbol-to-constellation mapping in the receiver is avoided using differential encoding in the transmitter. The part of the block diagram after the Demodulator block is used to extract the header and payload from the packets and then to check the CRC code to detect the correctly received bytes to be stored in an external file.

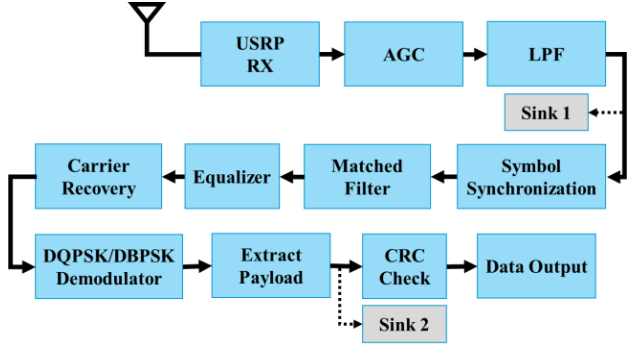

**Fig. 8**. The DBPSK and DQPSK receiver block diagram.

### C. Benchmark Test Module

The last step is measuring the performance of the developed communication system in terms of various metrics, including SNR, PER, PLR, throughput, latency, and communication distance. These metrics help examine the performance of wireless communication systems, which measure the effect of various impairments in the communication channel, such as noise, fading, interference, and attenuation. GNU Radio does not provide a tool to measure these performance metrics directly, so we implement a benchmark module to do this task as shown in Fig. 9. The Benchmark Test block we developed is a new contribution to the GNU Radio platform that can be used to evaluate the performance of various communication systems with realistic over-the-air transmission. This flexible tool can work with various communication systems, SDR hardware, communication protocols, and signal processing techniques using the CRC code in the transmitted packets. It relies only on radio signal and data packets and thus can be easily added and connected to other components in the GUI

GRC flowgraph.

This module has been uploaded to the GitHub online platform with usage and installation instructions to make it freely available to the GNU Radio community [31]. The benchmark module is developed using an out-of-tree (OOT) Python module. The core calculation process in the benchmark module is mainly implemented through two flow paths: a path for measuring the SNR and a path for measuring other metrics such as error rate and throughput. The benchmark module has two inputs connected via the corresponding sinks in the receivers shown in Figs. 7 and 8. The first input should be taken directly from the SDR hardware before demodulation via Sink 1, and the second input should be taken after demodulation before the CRC Check via Sink 2. These sinks are connected to the corresponding data source (Data 1 and Data 2) blocks in the benchmark module, as shown in Fig. 9. Thus, the complex signal received from the SDR hardware (e.g., the USRP) is first fed into the Fast-Fourier Transfer (FFT) block to be converted to the frequency domain. Then, the FFT samples are fed into the Benchmark Test block through the "in_sig" port to be used in the SNR calculation.

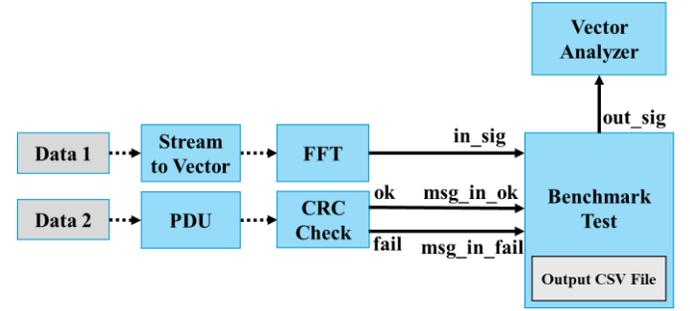

**Fig. 9**. Benchmark module block diagram.

In the Benchmark Test block, we first calculate the average power of each FFT bin over several iterations by taking the squared magnitude of the complex FFT bins normalized to the FFT vector size as given in (1). The number of iterations is adjusted using the parameter "No. of FFT iterations", as demonstrated in [31]. This average process can help in getting more accurate and reliable power measurements.

$$P_k = \sum_{i=1}^{M} \frac{|y_i(k)|^2}{M\,N^2} \tag{1}$$

where $P_k$ is the average power in watts for the $k^{th}$ FFT bin, $y_i(k)$ is the magnitude of the $k^{th}$ FFT bin at the $i^{th}$ iteration, N is the FFT vector size, and M is the number of iterations.

Now, the relative average power of each FFT bin is converted to dB and sent to the output port "out_sig" to visualize the averaged received signal using the vector analyzer (Vector Sink) block after adjusting its vector size parameter to the "Vector_Size" value, see [31].

To calculate the SNR, we first compute the channel power by summing the power of the FFT bins within the channel bandwidth using (2), which is equivalent to integrating the FFT bins over the channel bandwidth. This calculation process starts when the gateway starts receiving the data packets in order to get the relative received signal power.

$$P_s = \sum_{k=N_{first}}^{N_{last}} P_k \qquad (2)$$

where $N_{first}$ and $N_{last}$ are the indices of the first and last bins within the channel bandwidth, which make a vector of size N.

At the end of the transmission process, the benchmark algorithm starts measuring the relative noise power ($P_n$) for a specific duration of time following the same procedure of measuring the signal power described above. Then, the SNR is calculated in dB using the following equation:

$$\text{SNR}_{dB} = 10 \log_{10}\left(\frac{P_s - P_n}{P_n}\right) \qquad (3)$$

To calculate the system performance in terms of PER, PLR throughput, and latency, the received data stream after demodulation and payload extraction (but before performing the CRC check) should be converted to protocol data unit (PDU) messages to be compatible with the next block. Thus, the data stream taken from Sink 2 is first converted to PDU messages using the PDU block (i.e., Tagged Stream to PDU block). The PDU messages are passed into the CRC Check block to check whether the received packets are correct (ok) or incorrect (fail). The output ports of the CRC Check block are connected to the corresponding input ports on the Benchmark Test block, namely, "msg_in_ok" and "msg_in_fail" input ports. The PDU is sent via either the ok or fail output ports on the CRC Check block depending on the check result.

In the Benchmark Test block, two counters count the number of the correct and incorrect received packets, i.e., $N_{ok}$ and $N_{fail}$, respectively. We can also specify the number of discarded packets in this block to exclude those packets from the calculation. The PER is calculated as a ratio of the number of packets received incorrectly ($N_{fail}$) to the total number of transmitted packets ($N_{tx}$) [32]. A packet is considered incorrect if it contains at least one erroneous bit. The PLR is calculated as a ratio of the number of lost packets in the channel to the total number of transmitted packets. PER and PLR are often expressed as percentages or fractions. In general, PER and PLR ratios of less than 1% are acceptable for most applications.

$$PER = \frac{N_{fail}}{N_{tx}} \qquad (4)$$

$$PLR = \frac{N_{tx} - N_{ok}}{N_{tx}} \qquad (5)$$

In addition, we use time counters to measure the total transmission and delivery time to calculate the throughput and latency. Throughput (R) measures the number of packets successfully received per unit of time. Therefore, throughput is calculated in Kbps as follows.

$$R = \frac{8 \, N_{ok} \, L_p}{10^3 \, T} \qquad (6)$$

where $L_p$ is the packet length in bytes, and T is the total data delivery time in seconds. The latency (packet delivery time) is calculated by dividing the total time duration by the total number of packets received.

Finally, the Benchmark Test block exports the measured metrics to an external CSV file which includes the measurement time/date, SNR, $N_{fail}$, $N_{ok}$, PER, PLR, T, latency, R, received data size, and the corresponding configuration parameters. Also, a summary of the output data is displayed in the GRC console window. For more details on how to use this benchmark module, the reader can refer to the GitHub repository of this module [31].

## IV. RESULTS AND DISCUSSION

In this section, we evaluate the performance and feasibility of the developed narrowband SDR-based communication system in the TVWS band. We implemented the measurement testbed following the FCC regulation rules for operating narrowband devices in the TVWS band, as discussed in Section II. The system performance is evaluated using communication metrics, including PER, PLR, SNR, latency, and throughput. The measurement testbed is built using USRP SDR devices, consisting of a node (acts as an IoT node) and a BS (acts as an IoT gateway). IoT devices usually require higher uplink data traffic than downlink; thus, uplink is more critical in IoT applications. Therefore, we focus here on evaluating the performance and feasibility of the uplink path from the node to the BS. The outdoor measurement experiment was conducted at different node locations which have NLOS paths with the BS, as shown in Fig. 4 (b). The system parameters used in the outdoor measurement campaign are listed in Table III.

TABLE III
THE SYSTEM PARAMETERS USED IN THE OUTDOOR MEASUREMENT CAMPAIGN.

| Parameter | Value | Unit | Description |
|-----------|-------|------|-------------|
| $D_{TX}$ | 364 | KB | Size of the transmitted data file |
| $L_P$ | 500 | Bytes | Packet length |
| $h_n$ | 2 | m | Node height above ground level |
| $h_B$ | 12 | m | BS height above ground level |
| $G_{TX}$ | 5 | dBi | Transmitter antenna gain |
| $G_{RX}$ | 5 | dBi | Receiver antenna gain |
| $f_A$ | 470 - 614 | MHz | Antenna frequency range |
| $f_c$ | 600 | MHz | Center frequency |
| $P_{TX}$ | 12.6 | dBm | Transmit power |
| EIRP | 17.6 | dB | Effective isotropic radiated power |
| BW | 100 | kHz | Channel bandwidth |

The PER (%), PLR (%), and throughput values were obtained using various modulation techniques, i.e., DQPSK, DBPSK, GFSK, and GMSK, at different node locations as listed in Tables IV and V. We use transmit power of 12.6 dBm with 5 dBi antenna gain (i.e., EIRP is 17.6 dB). These metrics are obtained by taking the average values of multiple measurement trials. The PER and PLR values are obtained based on the measured number of correct and erroneous received packets using (4) and (5), respectively. In addition, the PER and PLR are plotted by taking the average value over various node locations, as shown in Fig. 10.

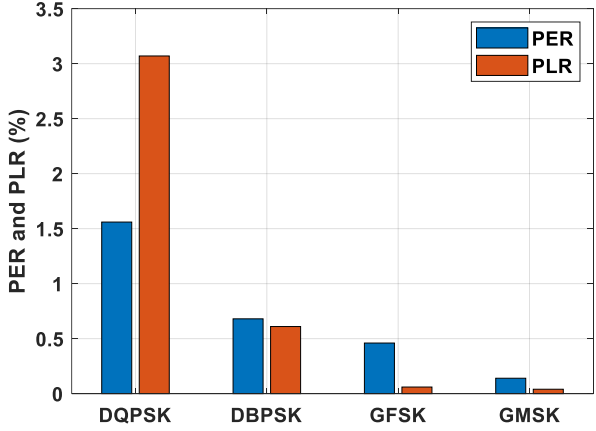

**Fig. 10**. The PER (%) and PLR (%) of various modulation schemes averaged over various node locations.

| Node | PER/PLR | DQPSK | DBPSK | GFSK | GMSK | SNR (dB) |
|------|---------|-------|-------|------|------|----------|
| A | PER (%) | 0.96 | 1.89 | 0.37 | 0.18 | 23.24 |
|   | PLR (%) | 0.42 | 0.99 | 0.07 | 0.15 | |
| B | PER (%) | 0.70 | 0.03 | 0.08 | 0.05 | 25.97 |
|   | PLR (%) | 6.01 | 0.32 | 0.04 | 0.06 | |
| C | PER (%) | 2.12 | 0.00 | 0.00 | 0.11 | 26.25 |
|   | PLR (%) | 3.90 | 0.25 | 0.12 | 0.00 | |
| D | PER (%) | 1.15 | 0.22 | 0.00 | 0.05 | 25.49 |
|   | PLR (%) | 0.63 | 0.24 | 0.06 | 0.03 | |
| E | PER (%) | 2.16 | 0.71 | 0.07 | 0.09 | 21.88 |
|   | PLR (%) | 0.73 | 1.12 | 0.07 | 0.06 | |
| F | PER (%) | 2.30 | 1.20 | 2.26 | 0.36 | 17.35 |
|   | PLR (%) | 6.75 | 0.75 | 0.02 | 0.00 | |
| Avg. | PER (%) | 1.56 | **0.68** | **0.46** | **0.14** | - |
|   | PLR (%) | 3.07 | **0.61** | **0.06** | **0.04** | - |

In general, the results indicate that the DQPSK modulation fails to satisfy the acceptable PER and PLR values (i.e., > 1%) for most node locations. Also, DQPSK provides the worst performance compared to other modulation schemes, with a PER of 1.56% and PLR of 3.07% on average. However, the DBPSK, GFSK, and GMSK techniques show acceptable performance with PER and PLR of less than 1% for most node locations. We can observe that the GMSK and GFSK modulation schemes provide the best performance with a PER of (0.14% and 0.46%) and PLR of (0.04% and 0.06%), respectively. The GMSK is widely used in wireless communication systems that require low power consumption over limited channel bandwidth. The developed narrowband communication system can successfully transmit packets with a communication distance of up to 3 km, i.e., the node-A location. This is the maximum distance from the BS that we can reach on campus. It is worth mentioning here that we have obtained permission from the CST to use the television broadcast spectrum for transmission within the campus. However, the developed system is expected to operate over longer distances, particularly for near-LOS and LOS scenarios and/or using higher antenna heights.

Throughput is calculated from the measurement data using (6), and the achieved throughputs for different modulation techniques at various node locations are listed in Table V. The measurements show that the developed narrowband SDR-based TVWS system can achieve throughput ranging between 94-97 Kbps on average for the node locations under analysis. The achieved throughput is considered satisfactory and sufficient for most IoT applications, considering that we are using a narrowband channel of 100 kHz. The total time taken to send and receive the text file (364 KB) using a packet length of 500 bytes is about 29 seconds on average. Thus, the packet delivery time (latency) is approximately equal to 40ms.

| Node | DQPSK | DBPSK | GFSK | GMSK |
|------|-------|-------|------|------|
| A | 96.74 | 95.44 | 97.38 | 97.53 |
| B | 96.96 | 97.73 | 97.68 | 97.71 |
| C | 92.42 | 97.76 | 97.76 | 97.66 |
| D | 96.53 | 97.53 | 97.76 | 97.72 |
| E | 95.49 | 96.62 | 97.69 | 97.63 |
| F | 90.14 | 96.45 | 95.52 | 97.24 |
| Avg. | **94.71** | **96.92** | **97.29** | **97.1** |

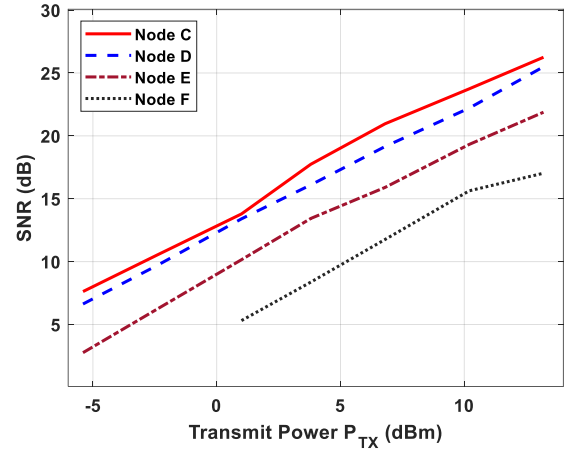

**Fig. 11**. The SNR versus the transmit power for various node locations.

We also show the impact of the transmit power value on the performance metrics for some propagation paths between the BS and the nodes. The SNR is calculated based on the relative measured value of the received signal and noise power using (1)-(3). As expected, Fig. 11 shows a linear increasing trend of the SNR with transmitting power for various nodes. We can observe that the communication link between BS and node-F has the lowest SNR, while the link between BS and node-C has the highest SNR. To explain why this happens, we did a local survey for these paths, and we also used both shuttle radar topography mission (SRTM) 1 arc-second digital elevation model (DEM) and OpenStreetMap (OSM) in MATLAB to investigate the propagation path. We found that the radio signal encounters many houses and walls in the path between the BS and the node-F while the path between the BS and the node-C encounters few houses, which explains these results.

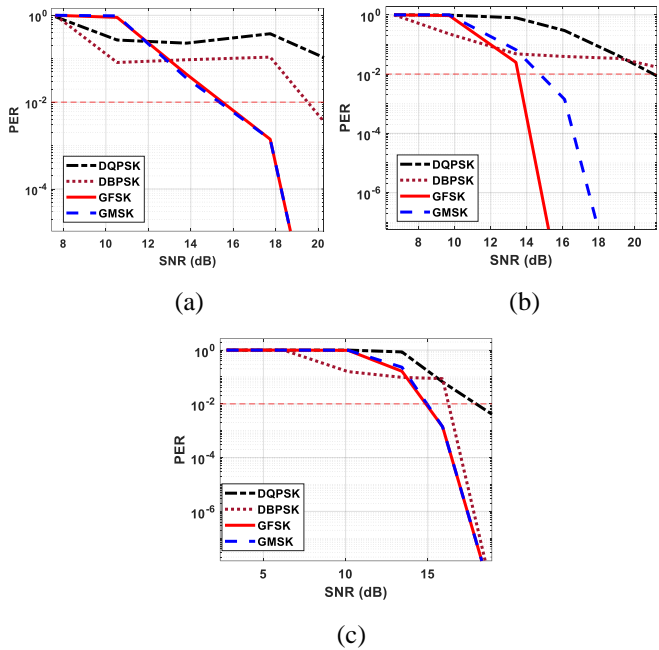

**Fig. 12**. The PER versus SNR for various modulation techniques for the following node locations: (a) Node-C, (b) Node-D, and (c) Node-E.

The PER and PLR as a function of SNR for various modulation techniques are plotted in Figs. (12)-(13) for nodes C, D, and E. We generally observe that the PER and PLR decrease as the SNR increases, which is an expected behaviour. The PER and PLR of DQPSK are observed to be greater compared to other modulation schemes. This is because the constellation points are closer to each other for higher-order modulation than lower-order modulation schemes. The theoretical assumption that BPSK and QPSK have the same error rate does not apply here because the system may have phase noise for various reasons, and the channel may not be modelled as Additive White Gaussian noise (AWGN).

In addition, DQPSK requires higher transmit power to achieve the required error performance. DQPSK and DBPSK show larger PER and PLR values than GMSK and GFSK modulation schemes. This could be due to an imperfection in the carrier synchronization and equalization processes at the BS (i.e., the receiving USRP). We noticed that after the transmission process started by the node, it took a few seconds for the BS to lock onto the correct carrier, causing error packets during this time. The PSK modulation is more sensitive to frequency and phase offset. Thus, precise carrier synchronization between transmitter and receiver is required, which increases the design complexity. For example, the minimum SNR values (SNR$_{min}$) needed to achieve 1% PER for node-C are about 16 dB for GMSK/GFSK, 20 dB for DBPSK, and a larger SNR$_{min}$ value is needed for DQPSK, which does not appear in this figure. For node D, SNR$_{min}$ is about 14-15 dB for GFSK/GMSK and 21 dB for DBPSK/DQPSK. Moreover, for node E, SNR$_{min}$ is about 15 dB for GFSK/GMSK, 16 dB for DBPSK, and 18 dB for DQPSK. The measured data show that GMSK and GFSK provide better performance with less energy requirement (i.e., less SNR$_{min}$) than DBPSK and DQPSK modulation schemes.

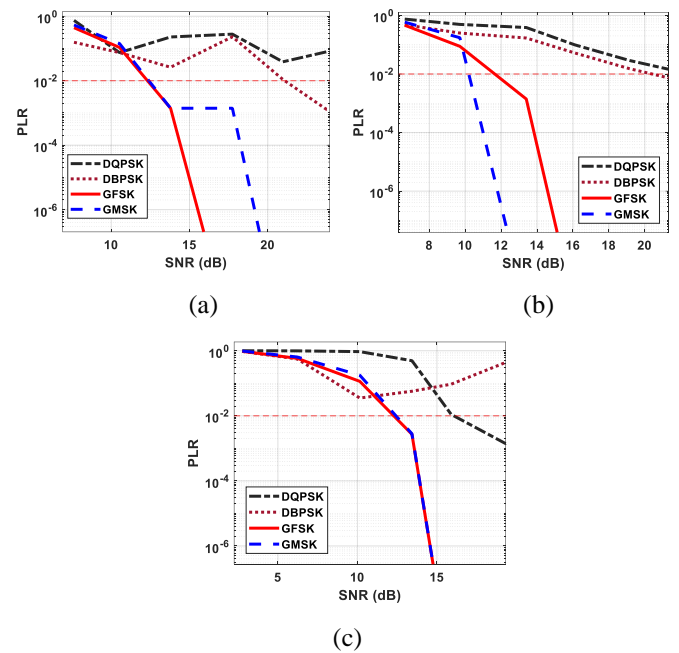

**Fig. 13**. The PLR versus SNR for various modulation techniques for the following node locations: (a) Node-C, (b) Node-D, and (c) Node-E.

In addition, the results show that throughput increases with the SNR to finally reach the maximum limit of about 97 Kbps, as shown in Fig. 14. The maximum throughput achieved using GMSK and GFSK requires a minimum SNR of about 16 dB. In contrast, DBPSK and DQPSK require SNR larger than 20 dB. In general, the DQPSK modulation scheme shows higher throughput than DBPSK because the bandwidth efficiency of DQPSK is twice that of DBPSK, where two bits per symbol are transmitted instead of one bit per symbol. The achieved throughput is sufficient for most IoT applications that require low data rates. NB-WSD is allowed to transmit for no more than 36 seconds per hour, so this data rate is sufficient to send various data such as sensor readings and control messages.

A trade-off between complexity and performance should be considered in designing IoT wireless communication systems, including LPWANs. We found that GMSK and GFSK modulation schemes provide the best performance in terms of PER, PLR, and throughput compared to DBPSK and DQPSK.

The GMSK and GFSK have better bandwidth efficiency, and less interference compared to DBPSK and DQPSK due to the use of a smoother transition pulse by the Gaussian filter, which reduces the out-of-band radiation compared to the rectangular pulse used in the PSK modulation schemes. In addition, the GFSK and GMSK require lower SNR to achieve an acceptable level of error compared to DQPSK and DBPSK. It is worth noting that energy consumption in the data transmission process is one of the crucial requirements for designing IoT systems. In addition, the design complexity of the modulation and demodulation in DQPSK and DBPSK systems is greater than in GMSK and GFSK systems. Therefore, we believe that GMSK and GFSK can provide satisfactory performance at a lower cost and with less design complexity for narrowband TVWS-based wireless communication systems. In this work, the testbed is implemented as a prototype to examine the feasibility and performance of the narrowband communication system in the TVWS band for IoT applications using various modulation techniques. However, practical deployment of narrowband IoT nodes requires redesigning this system into small battery-powered units instead of using SDR USRP devices.

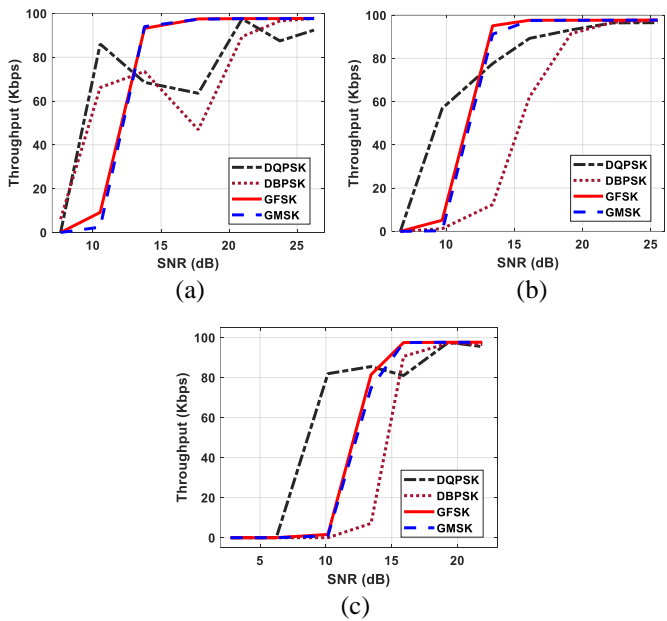

**Fig. 14**. Throughput versus SNR for various modulation techniques at the following node locations: (a) Node-C, (b) Node-D, and (c) Node-E.

## V. CONCLUSION

In this paper, we examine the feasibility and performance of narrowband data communication in the TVWS band for IoT applications. We implement an SDR-based measurement testbed consisting of a node communicating with a BS over the uplink path using various modulation schemes, including DQPSK, DBPSK, GFSK, and GMSK. The developed testbed is used for conducting outdoor measurements at various locations on the KAUST campus over NLOS propagation paths to evaluate the system performance in terms of PER, PLR, latency, throughput, SNR, and communication distance. In addition, we have developed a flexible and easy-to-use benchmark module which provides a powerful tool to evaluate the over-the-air performance of various SDR-based communication systems in GNU Radio. The results show the efficiency and capability of the proposed system in providing a reliable communication link with sufficient throughput and acceptable PER and PLR for most low-data rate IoT applications. This system can provide a throughput of up to 97 Kbps with PER and PLR of less than 1% over NLOS paths up to 3 km. This system is expected to cover longer distances via LOS and near-LOS propagation paths and/or using higher antenna heights. The DQPSK modulation technique performs worse with PER and PLR of more than 1%. On the other hand, GFSK and GMSK techniques show acceptable PER and PLR of less than 1%. In addition, the results show that GMSK and GFSK require lower $SNR_{min}$ than DQPSK and DBPSK to achieve PER and PLR below 1%. The PSK modulation seems to be more sensitive to frequency offset and phase noise which require precise carrier synchronization. It is worth mentioning that the system designer should consider a trade-off between the system's complexity, performance, and energy consumption. This work can provide insight into the feasibility and efficiency of narrowband wireless communication in the TVWS band for IoT applications using different modulation techniques. Therefore, this work can help in the physical layer design of small battery-powered narrowband radio transceivers operating in the TVWS band. Furthermore, we believe that the developed benchmark module and testbed can provide great assistance to the researchers and engineers in the GNU Radio community in evaluating the performance of various SDR-based communication systems. In future work, we will expand this study to rural and urban areas to examine more channel conditions.

## ACKNOWLEDGMENT

We would like to thank the Communications, Space, and Technology Commission (CST) in Saudi Arabia for granting us permission to conduct this experiment using the TV spectrum on the KAUST campus.

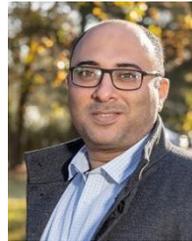

**Muneer M. Al-ZuBi** (Member, IEEE) received his Ph.D. degree in engineering from the University of Technology Sydney (UTS), Sydney, Australia, in 2020. He worked as a Research Associate with the Center of Excellence for Innovative Projects, Jordan University of Science and Technology (JUST), from 2020 to 2021. From 2021 to 2022, he was a postdoctoral researcher with the Department of Engineering, University of Luxembourg, Luxembourg, and also, he was a remote visiting scholar with the School of Electrical and Data Engineering, UTS. He is currently a Postdoctoral Researcher at the Communication Theory Lab (CTL), King Abdullah University of Science and Technology, Saudi Arabia. His research interests lie in the areas of wireless communication, EM wave propagation, and molecular communication.

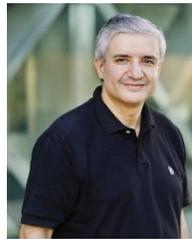

**Mohamed-Slim Alouini** (Fellow, IEEE) received his Ph.D. degree in electrical engineering from the California Institute of Technology (Caltech), Pasadena, CA, USA, in 1998. He served as a faculty member at the University of Minnesota, Minneapolis, MN, USA, then at Texas A&M University at Qatar, Education City, Doha, Qatar before joining King Abdullah University of Science and Technology (KAUST), Thuwal, Makkah Province, Saudi Arabia as a professor of electrical engineering in 2009. His current research interests include the modeling, design, and performance analysis of wireless communication systems.